\documentclass{article}
\usepackage{epic}
\usepackage{epsfig}
\usepackage{subfigure}
\usepackage{curves}
\usepackage{amsmath}
\usepackage{amssymb}
\usepackage{rotating}
\begin{document}
\newcommand{\abs}[1]{\lvert#1\rvert}
\newcommand{\DoT}[1]{\begin{turn}{-180}\raisebox{-1.7ex}{#1}\end{turn}}
\title{{\bf Structures generated by the Klein-like 
topology of space}\thanks{Poster at the 
TH2002 conference,  Paris, 22-26 July 2002}}
\author{Vladimir Yershov \\
\vspace{3.0\baselineskip}                                               
{\small \it vny@mssl.ucl.ac.uk} \\
\vspace{3.0\baselineskip}                                               
{\small \it Mullard Space Science Laboratory (University College London),} \\
{\small \it Holmbury St.Mary, Dorking RH5 6NT, United Kingdom} \\    
}     

\date{}
\maketitle

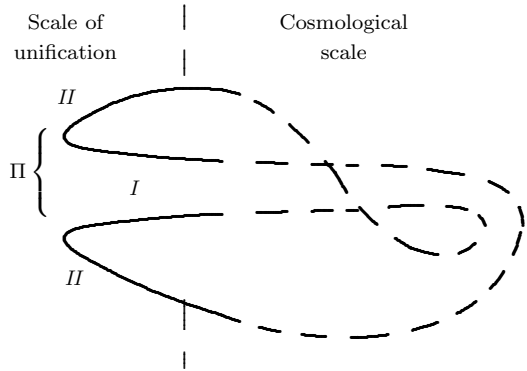
\begin{figure}[htb]
\centering
\setlength{\unitlength}{1mm}
\begin{picture}(70,50)(-2,-2)
 \linethickness{1pt}
 \qbezier(30,28.5)(20,29)(13,30)
 \qbezier(13,30)(5,31)(13,35)
 \qbezier(13,35)(20,39)(30,38)
 \qbezier(30,21.5)(20,21)(13,20)
 \qbezier(13,20)(5,19)(13,15)
 \qbezier(13,15)(20,11)(30,8)
 \curvedashes[1mm]{0,3,3}
 \qbezier(48,28)(45,28)(30,28.5)
 \qbezier(48,22)(45,22)(30,21.5)
 \qbezier(30,38)(37,37)(45,27)
 \qbezier(45,27)(48,21)(55,17)
 \qbezier(55,17)(63,14)(65,20)
 \qbezier(65,20)(64,24)(48,22)
 \qbezier(30,8)(37,5)(48,5)
 \qbezier(48,5)(67,6)(70,20)
 \qbezier(70,20)(70,26)(60,27)
 \qbezier(60,27)(55,28)(48,28)
 \linethickness{0.5pt}
 \qbezier(25,1)(25,3)(25,10)
 \qbezier(25,40)(25,45)(25,50)
 \curvedashes{}
 \thinlines
 \put(5,33){
 \makebox(0,0)[t]{\footnotesize{$\Pi \begin{cases}
 \text{\hspace{0.01cm}} \\
 \text{\hspace{0.01cm}} \\
 \text{\hspace{0.01cm}}
 \end{cases}$}}}
 \put(17,26){
 \makebox(0,0)[t]{\footnotesize{\bf \it  I}}}
 \put(8,38){
 \makebox(0,0)[t]{\footnotesize{\bf \it II}}}
 \put(9,14){
 \makebox(0,0)[t]{\footnotesize{\bf \it II}}}
 \put(8,48){
 \makebox(0,0)[t]{\footnotesize{Scale of}}}
 \put(8,44){
 \makebox(0,0)[t]{\footnotesize{unification}}}
 \put(45,48){
 \makebox(0,0)[t]{\footnotesize{Cosmological}}}
 \put(45,44){
 \makebox(0,0)[t]{\footnotesize{scale}}}
\end{picture}
\caption{{\small One-dimensional scheme of the Klein-bottle.
The ``inner'' ({\it I}) and the ``outer'' ({\it II})
hyper-surfaces are unified through the region $\Pi$ 
regarded here as a primitive particle}}
 \label{fig:klein1}
\end{figure}

We hypothesise that properties of space could underly some patterns observed 
in nature. Let us explore the possibility that the observed variety
of matter particles and the pattern of their properties are determined by 
the non-orientability of the manifold representing spacetime. 
That is, let us regard space as a moving medium with the topology
of a 3-Klein bottle, whose motion is parameterised with the 
time coordinate.
The ``throat'' of the Klein bottle (region $\Pi$ in Fig.\ref{fig:klein1})
could be viewed as a dislocation of space with vanishing radius, so that the
unification of the ``inner'' (I) and ``outer'' (II) hyper-surfaces
of the Klein bottle occur on a very small scale.

\vspace{0.1cm}

\begin{figure}[htb]
\centering
{\epsfig{figure=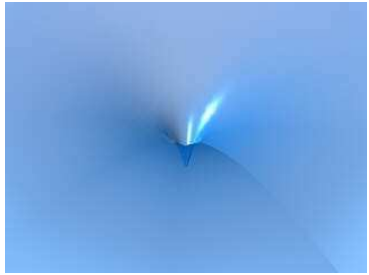, width=.3\textwidth}}
\caption{Two-dimensional visualisation of the preon $\Pi$.}
 \label{fig:klein2}
\end{figure}

The resulting structure resembles a microscopic black hole with its
inner and outer hyperspaces identified. Typically, this could be 
visualised (as a point-like source) by reducing dimensionality and 
using a two-dimensional surface, as shown in Fig.\ref{fig:klein2}.

We shall regard this structure as a primitive particle (hereafter 
called ``preon'') with its field extending to infinity. 
Since the  inner space of this particle is identified with the rest
of the system, such a model automatically incorporates Mach's principle.
This also means that preon's properties are determined by the 
parameters of this system (the universe) and that in multi-preon systems 
preons must be identical to each other, save their differences due to 
the torsional degrees of freedom of the medium.

In such a system different preons would interact with one another
through pressure and tension of the moving medium. The torsional degrees of 
freedom would give rise to the 
SU(3)/SU(1)-symmetry of the fields (due to the tridimensionality of space). 
By analogy with quantum chromodynamics (dealing with these types of symmetry), 
we can endue preons with tripolar  
charges (colours), which are usually labelled as $red$, 
$green$, and $blue$. For simplicity let us use unit values 
for these charges, as well as for the preons' masses.

Given two manifestations of space, we can resolve the preon's field
into two components, $\phi_s$ and $\phi_e$. To avoid singularities we shall assume 
that infinite energies are not accessible and hypothesise that the 
energy of both $\phi_s$ and $\phi_e$, after reaching the maximum,
decays to zero at the origin. The simplest form for the split
field that incorporates the requirements above is the following:
\begin{equation}
\begin{split}
& F=\phi_s+\phi_e, \\
& \phi_s=s\exp{(-\rho^{-1})}, \hspace{0.5cm} \phi_e=-\phi'_s(\rho),
\end{split}
\label{eq:basicfield}
\end{equation}
\noindent
where the signature $s=\pm 1$ indicates the sense of the interaction (attraction
or repulsion); the derivative of $\phi_s$ is taken with respect to the radial
coordinate $\rho$. Far from the source, the second component of the field $F$ mimics 
the Coulomb gauge, whereas the first component extends to infinity being
almost constant (similarly to the strong field).

To formalise the calculation of masses, we shall represent 
the preon discharge with the use of 
auxiliary  $3\times 3$ singular matrices ~$\Pi^i$ containing 
the following elements:
\begin{equation}
^{\pm}\pi^i_{jk}=\pm\delta^i_j (-1)^{\delta^k_j},
\label{eq:pmatrix}
\end{equation}
where $\delta^i_j$ is the Kronecker delta-function;
the $\pm$-signs correspond to the sign of the charge; 
and the index $i$ stands for the colour ($i=1,2,3$ or red, green and blue).
The diverging components of the field can be represented by 
reciprocal elements:
$\tilde{\pi}_{jk}=\pi_{jk}^{-1}$.
Then, we can define the preon's (unit) charges and masses 
by summation of these matrix elements:  
\begin{equation}
\begin{split}
q_{\Pi}&=\mathbf{u}^{\intercal}{\Pi}\mathbf{u} \hspace{0.1cm},
 \hspace{0.3cm} \tilde{q}_{\Pi}=\mathbf{u}^{\intercal} \tilde{\Pi}
\mathbf{u} \\
m_{\Pi}&=\abs{\mathbf{u}^{\intercal}{\Pi}\mathbf{u}} \hspace{0.1cm},
\hspace{0.3cm} \tilde{m}_{\Pi}=\abs{\mathbf{u}^{\intercal} \tilde{\Pi}
\mathbf{u}}
\end{split}
\label{eq:preonchargemass}
\end{equation}
($\mathbf{u}$ is the diagonal of a unit matrix; $\tilde{q}_\Pi$ 
and $\tilde{m}_\Pi$ diverge).   
 Let us assume that the field $F(\rho)$ does not act 
instantaneously at a distance. Then, we can define the mass 
of a system containing, say, $N$ preons, as 
proportional to the number of these particles, 
wherever their field flow rates are not cancelled.
For this purpose, we shall regard  the total 
field flow rate, $v_N$, of such a system as a 
superposition of the individual volume flow rates 
of its $N$ components. Then, the net mass of the system can 
be calculated (to a first order of approximation) as the number
of particles, $N$, times the normalised to unity (Lorentz-additive)
field flow rate $v_N$:
\begin{equation}
m_N=\abs{N v_N}.
\label{eq:mass}
\end{equation}
\noindent
Here $v_N$ is computed recursively from:
\begin{equation}
v_i=\frac{q_i+v_{i-1}}{1+\abs{q_i v_{i-1}}}, 
\label{eq:flowrate}
\end{equation}
with $i=2, \dots\,,N$ and putting $v_1=q_1$.
The normalisation condition (\ref{eq:flowrate}) 
expresses the common fact that the superposition
flow rate of, say, 
two antiparallel flows ($\uparrow \downarrow$) with equal
rate magnitudes 
$\abs{\mathbf{v}_\uparrow}=\abs{\mathbf{v}_\downarrow}=v$ vanishes 
($v_{\uparrow \downarrow}=0$), whereas, in the case of parallel
flows ($\uparrow \uparrow$) it cannot exceed the magnitudes of   
the individual flow rates ($v_{\uparrow \uparrow} \leq v$).

Then, when two unlike-charged preons combine (say, red and antigreen),
the magnitudes of their oppositely directed flow rates cancel each other
(resulting in a neutral system). The corresponding acceleration 
also vanishes, which is implicit in (\ref{eq:mass}).
This formula implies the complete cancellation of masses in the
system with vanishing electric fields (converted into the binding
energy of the system), but this is only an approximation 
because in our case the preons are separated by a distance of equilibrium, 
whereas the complete cancellation of flows is possible only when the flow
source centres coincide.

Making use of the known pattern of attraction and repulsion
between colour charges 
(two like-charged but unlike-coloured particles are
attracted, otherwise they repel) we can write the signature $s_{ik}$
of the chromoelectric interaction between two preons with 
colours $i$ and $k$ as

\begin{equation}
s_{ik}=-\mathbf{u}^\intercal{\Pi}^i {\Pi}^k\mathbf{u}.
\label{eq:seforce}
\end{equation}

\begin{figure}[htb]
\centering
\setlength{\unitlength}{0.8mm}
\begin{picture}(100,50)(-10,10)
 \curvedashes[1mm]{0,1,2}
 \put(20,40){\arc(0,4){360}}
 \put(18,40){\line(1,0){4}}
 \put(20,38){\line(0,1){4}}
 \put(20,30){\arc(0,4){360}}
 \put(18,30){\line(1,0){4}}
 \put(20,28){\line(0,1){4}}
 \put(90,40){\arc(0,4){360}}
 \put(88,40){\line(1,0){4}}
 \put(90,38){\line(0,1){4}}
 \put(90,30){\arc(0,4){360}}
 \put(88,30){\line(1,0){4}}
 \put(90,28){\line(0,1){4}}
 \curvedashes{}
 \put(0,40){\arc(0,4){360}}
 \put(-2,40){\line(1,0){4}}
 \put(0,38){\line(0,1){4}}
 \put(0,30){\arc(0,4){360}}
 \put(-2,30){\line(1,0){4}}
 \put(0,20){\arc(0,4){360}}
 \put(-2,20){\line(1,0){4}}
 \put(0,18){\line(0,1){4}}
 \put(0,10){\arc(0,4){360}}
 \put(-2,10){\line(1,0){4}}
 \put(70,40){\arc(0,4){360}}
 \put(68,40){\line(1,0){4}}
 \put(70,38){\line(0,1){4}}
 \put(70,30){\arc(0,4){360}}
 \put(68,30){\line(1,0){4}}
 \put(70,20){\arc(0,4){360}}
 \put(68,20){\line(1,0){4}}
 \put(70,18){\line(0,1){4}}
 \put(70,10){\arc(0,4){360}}
 \put(68,10){\line(1,0){4}}
 \put(20,20){\arc(0,4){360}}
 \put(18,20){\line(1,0){4}}
 \put(20,18){\line(0,1){4}}
 \put(20,10){\arc(0,4){360}}
 \put(18,10){\line(1,0){4}}
 \put(20,8){\line(0,1){4}}
 \put(90,20){\arc(0,4){360}}
 \put(88,20){\line(1,0){4}}
 \put(90,18){\line(0,1){4}}
 \put(90,10){\arc(0,4){360}}
 \put(88,10){\line(1,0){4}}
 \put(90,8){\line(0,1){4}}
 \put(10,50){ \makebox(0,0)[t]{Strong interaction}}
 \put(80,50){ \makebox(0,0)[t]{Electric interaction}}
 \put(-8,29){$\leftarrow$}
 \put(-8,19){$\leftarrow$}
 \put(12,39){$\leftarrow$}
 \put(12,9){$\leftarrow$}
 \put(62,39){$\leftarrow$}
 \put(62,9){$\leftarrow$}
 \put(82,29){$\leftarrow$}
 \put(82,19){$\leftarrow$}
 \put(4,39){$\rightarrow$}
 \put(24,29){$\rightarrow$}
 \put(24,19){$\rightarrow$}
 \put(4,9){$\rightarrow$}
 \put(94,39){$\rightarrow$}
 \put(74,29){$\rightarrow$}
 \put(74,19){$\rightarrow$}
 \put(94,9){$\rightarrow$}
\end{picture}
\vspace{0.1cm}
\caption{{\small Pattern of attraction and repulsion between preons
of different colour and charge polarities for two complementary
manifestations of the chromoelectric interaction.
Dashed and solid circles denote opposite (unlike) colours; 
charge polarities are indicated with the ``$+$'' and ``$-$'' signs.}}
 \label{fig:force}
\end{figure}
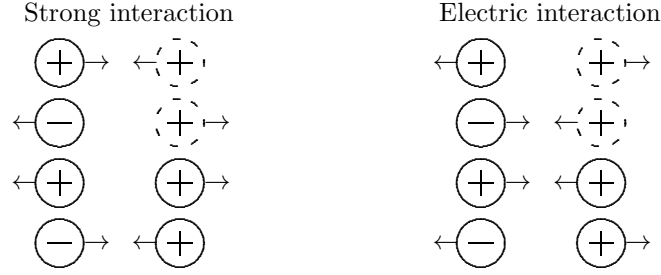

\noindent
The corresponding pattern is shown in Fig.\ref{fig:force}.
It implies that the preons of
different colours and charge polarities will cohere in structures, 
the simplest of which will be the charged
and neutral colour-doublets (dipoles): 

\begin{equation*}
\varrho^\pm_{ik}= \mathbf{\Pi}^i+\mathbf{\Pi}^k, 
\end{equation*}
\begin{equation*}
\hspace{2.5cm}g^0_{ik}= \mathbf{\Pi}^i+\overline{\mathbf{\Pi}}^k, \hspace{0.6cm} i,k=1,2,3
\end{equation*}

\noindent    
as well as the charged tripoles, denoted here  
as {\sf Y} and $\overline{\sf Y}$ (for the opposite polarities): 

\begin{equation*}
{\sf Y}=\sum_{i=1}^3{\mathbf{\Pi}^i} \hspace{0.4cm} {\rm or} \hspace{0.4cm}
\overline{{\sf Y}}=\sum_{i=1}^3{\overline{\mathbf{\Pi}}^i}.
\end{equation*}

\noindent
According to (\ref{eq:preonchargemass}), $q(\varrho_{ik})=\pm 2$, 
$m(\varrho_{ik})=2$, $q(g_{ik})=0$,
$m(g_{ik})=0$, $q({\sf Y})= \pm 3$, $m({\sf Y})= 3$.
The tripole ({\sf Y}-particle) will be colourless at infinity but 
colour-polarised nearby, which means that tripoles could
combine in strings (pole-to-pole to each other) due to their residual chromaticism.

\begin{figure}[htb]
\centering \epsfysize=9cm
\includegraphics[scale=0.30]{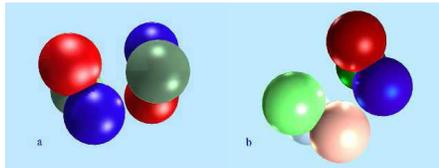}
\caption{{\small (a):Equilibrium configurations of two like-charged 
tripoles (rotated by $180^\circ$);  and 
(b): two unlike-charged tripoles
rotated by $120^\circ$ with respect to one another.}}
\label{fig:twoy}
\end{figure}

Pairs of like-charged tripoles {\sf Y} will combine in short strings 
 {\sf Y} \DoT{\sf Y}\,, with the components $180^\circ$-rotated with respect to 
each other, Fig.\ref{fig:twoy} ($a$), whereas unlike-charged tripoles 
({\sf Y}$\overline{\sf Y}$) would combine rotated by $\pm 120^\circ$,   
  Fig.\ref{fig:twoy} ($b$).

\begin{figure}[htb]
\centering \epsfysize=9cm
\includegraphics[scale=0.40]{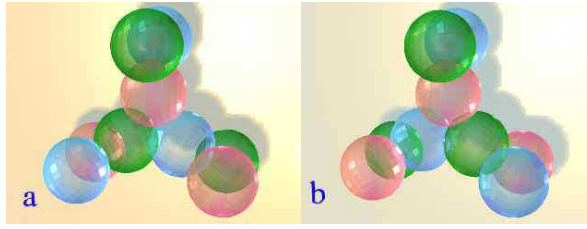}
\caption{{\small Three like-charged tripoles joined with their vertices 
directed towards the centre of the structure. 
The colour currents due to the motions of colour-charges along the loop  
appear as anticlockwise (a) or clockwise (b) helices.}}
\label{fig:electron}
\end{figure}

\begin{figure}[htb]
\centering \epsfysize=9cm
\includegraphics[scale=0.40]{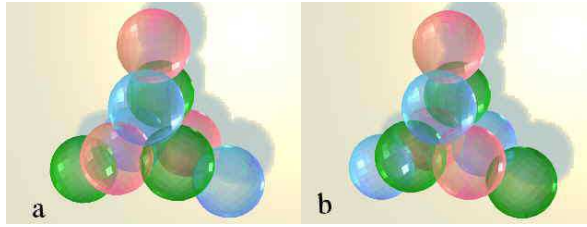}
\caption{{\small Three like-charged tripoles joined together with their 
vertices directed away from the centre of the structure.
The colour-currents along the loop  
appear as anticlockwise (a) or clockwise (b) helices.}}
\label{fig:electron2}
\end{figure}

The string of three like-charged tripoles will close in 
a loop (triplet) 3{\sf Y} or 3$\overline{\sf Y}$, denoted here as $e$
because, as we shall see, this structure by its properties can be identified 
with the electron. The triplet 3{\sf Y} is charged, with the charge $q_e=\pm 9$ and 
mass $m_e=9$ (expressed in units of preon's charge and mass).
The tripoles in this structure can be directed with 
their vertices towards or away from the centre 
of the loop, as shown in Fig.\ref{fig:electron} and Fig.\ref{fig:electron2}.
However, it is seen that these configurations correspond to two different 
phases of the same structure, since the tripoles here have a rotational 
degree of freedom (around their common ring-closed axis). 
At the same time, the tripoles
will orbit the centre of the structure moving along 
the ring-closed axis. The resulting currents have helical 
shapes with two possible helicity signs (clockwise or anticlockwise). 
These different helicities can be identified with two 
spins of the structure ($e_\uparrow$ and $e_\downarrow$).

\begin{figure}[htb]
\centering \epsfysize=9cm
\includegraphics[scale=0.6]{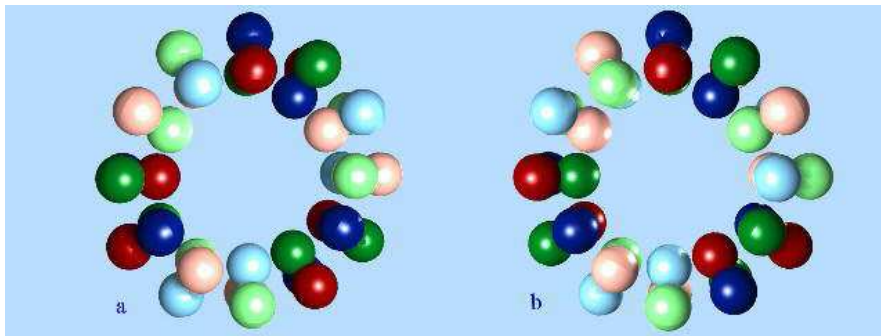}
\caption{{\small The clockwise (a) and anticlockwise
(b) helical currents can be identified with two opposite 
spins of the structure $6{\sf Y}\overline{\sf Y}$}}
\label{fig:twonu}
\end{figure}

\vspace{0.1cm}
The pairs of unlike-charged tripoles can form longer strings: 
{\sf Y}$\overline{\sf Y}$  \DoT{$\overline{\sf Y}$}\DoT{\sf Y}
  {\sf Y}$\overline{\sf Y}  \ldots$
with the pattern of colour charges repeating after each six 
consecutive {\sf Y}$\overline{\sf Y}$-pairs, which allows the closure of
such a string in a loop (shown in Fig.\ref{fig:twonu} and denoted here as 
$\nu_e=6{\sf Y}\overline{\sf Y}$). 
The structure $\nu_e$ (formed of 36 preons) is electrically neutral and 
has a vanishing mass, according to (\ref{eq:mass}), unless combined with 
a charged particle, say {\sf Y} or 3{\sf Y}, which would restore the entire
mass of the composite system.

The properties of the simplest preon structures 
are summarised in Table \ref{t:preonsummary}.
Most of these structures are open strings, which are supposed to be unstable.
The ring-closed structures, like $e$ and $\nu_e$, will be stable because their effective
potentials are minimised.

\begin{table}[htb]
\caption{Simple structures generated  by the 3-Klein bottle topology of space 
}
\label{t:preonsummary}
\begin{center}
\small
\begin{tabular}
{|c|c|r|r|r|} \hline
Structure     & Constituents       & Number of colour           & Charge~~~~~       & Mass~~~~~            \\
              & of the             & charges in the~~~~         & (in preon~~~~      & (in preon~~~~      \\
              & structure          & structure~~~~~~              & charge units)  & mass units)  \\ \hline
\multicolumn{5}{|c|} {The primitive particle (preon $\Pi$)} \\ \hline 
$\Pi$         & $1\Pi$                         &  1        & $+1$       &   1           \\ \hline
\multicolumn{5}{|c|} {First-order structures (combinations of preons} \\ \hline 
$\varrho$       & $2\Pi$                       &  2        & $+2$       &   2           \\ 
$g^0$         & $1\overline{\Pi}+1\Pi$         &  2        & $-1+1=0$   &   $\approx 0$           \\
{\sf Y}       & $3\Pi$                         &  3        & $+3$       &   3           \\  
\hline
\multicolumn{5}{|c|}{Second-order structures (combinations of tripoles {\sf Y})} \\
\hline 
$\delta^\pm$  & 2{\sf Y}                       &  6        & $+6$       &   6            \\ 
$\gamma^0$    & $1\overline{\sf Y}+1{\sf Y}$   &  6        & $-3+3=0$    &   $\approx 0$             \\ 
$e^-$         & $3\overline{\sf Y}$            &  9        & $-9$        &   9             \\ \hline
\multicolumn{5}{|c|}{Third-order structures}  \\
\hline
$2e^-$      & $3\overline{\sf Y}+3\overline{\sf Y}$ & $9+9=18$ &   $-18$       &  18 \\
$e^-e^+$    & $3\overline{\sf Y}+3{\sf Y}$     & $9+9=18$      & $-9+9=0$     &  $\approx 16$  \\
$\nu_e$     & $6\overline{\sf Y}{\sf Y}$       & $6\times(3+3)=36$  & $6\times(-3+3)=0$      &  $\approx 0$  \\
{\sf Y}$_1$ & $\nu$ \ \   + \ \  {\sf Y} & $36+3=39$           & $0-3=-3$   &  $36+3=39$ \\
$W^-$   & $\nu_e$ \ \ + \ \ $e^-$     & $36+9=45$ & $0-9=-9$        &  $36+9=45$ \\
$u$       & {\sf Y}$_1$ \ \ $\nu_e$ \ \ {\sf Y}$_1$ & $39+36+39=114$   & $+3+0+3=+6$        & $39+39=78$ \\
$\nu^0_\mu$   & {\sf Y}$_1$ \ \ $\nu_e$ \ \ $\overline{{\sf Y}}_1$ & $39+36+39=114$     & $-3+0+3=0$   &  $\approx 0$ \\
$d$       & $u$ \ \ + \ \ $W^-$ & $114+45=159$ & $+6-9=-3$ & $78+45=123$ \\
$\mu^-$     & \ $\nu_{\mu}$ \  + \ \ $W^-$ & $114+45=159$ & $0-9=-9$ & $\overline{48+39}^{\,*}=1872$ \\
\multicolumn{5}{|c|}{and so on ...}  \\     
\hline
\end{tabular}
\end{center}
{\footnotesize \vspace{-0.2cm}\hspace{11cm}*)\,two-component system }
\end{table}

\vspace{0.1cm}
The particles {\sf Y}, $e$, and $\nu_e$ can combine with each other
because of their residual chromaticism. The structure 
{\sf Y}$_1$={\sf Y}$+\nu_e$ will have mass and be 
charged, with the  charge $\pm 3$ units, corresponding to the charge of 
the {\sf Y}-particle,  and the mass of 39 preon mass units: 
$m_{{\sf Y}_1}=n_{\nu_e}+m_{\sf Y}=36+3$. 
The charge of the configuration $e+\nu_e$ will correspond to 
the charge of the triplet  \ $e$  ($\pm 9$ units). Its mass will be 45 units, 
if expressed in  preon's units of mass ($n_{\nu_e}+m_e=36+9$).

\vspace{0.1cm}
Like-charged particles {\sf Y}$_1$ of the same helicity signs 
would further combine (through an intermediate 
$\nu_e$-particle with the opposite helicity) forming
three-component strings. 
The string {\sf Y}$_1\,e$\,\,\DoT{\sf Y}$_1$ can be identified with 
the {\it u}-quark. Its charge will correspond to the charge of 
two {\sf Y}-particles ($q_u=\pm 6$) and its mass will roughly
be the sum of the masses of its two ${\sf Y}_1$-components:
$m_u=2\times 39=78$ (preon mass units). The positively charged $u$-quark
(78 preon mass units) would be able to combine with the negatively charged particle 
$e^-\nu_e$ (45 preon mass units) mass, forming the {\it d}-quark with a mass
of 123 preon mass units, $m_d=m_u+m_{e\nu_e}=78+45=123$, 
 and with the charge derived from the charges of its constituents 
$q_d=q_u+q_e=+6-9=-3$ units.

\vspace{0.1cm}
The process of structure formation involving the particles 
{\sf 3Y} with $6{\sf Y}\overline{\sf Y}$  ($e$ and $\nu_e$) will 
be helicity-dependent. 
The configuration of colour charges of the structure $6{\sf Y}{\overline{\sf Y}}_\uparrow$  
does not match that of the structure {\sf 3Y$_\downarrow$},
which would lead to the mutual repulsion of these particles.
Only the structures {\sf 3Y}$_\uparrow$ + $6{\sf Y}{\overline{\sf Y}}_\uparrow$ 
or {\sf 3Y}$_\downarrow$ + $6{\sf Y}{\overline{\sf Y}}_\downarrow$ can be formed
because the effective potential of these combinations implies attraction
between the components.
By contrast, if two structures of the same kind combine (e.g., $e$ with $e$ 
or $\nu_e$ with $\nu_e$), their helicity signs must 
be opposite to create an attractive force between the components of the pair.  
This coheres with and probably explains the Pauli exclusion principle, which
allows us to identify helicities of the structures in question with their 
spins. 
  
\vspace{0.1cm}
Divided by nine, the charge of the 3{\sf Y}-particle,
gives us the conventional unit charge of the electron.
Charges of the {\sf Y} and $\delta$-particles 
(fractions of the nine-unit electron charge)
correspond to the quark fractional charges.

\begin{figure}[htb]
\centering \epsfysize=9cm
\includegraphics[scale=0.52]{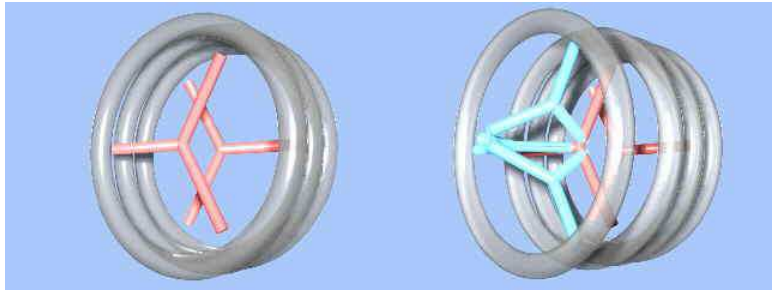}
\caption{{\small Schemes of the $u$ (left) and $d$ 
(right) quarks.}}
\label{fig:updown}
\end{figure}

\vspace{0.1cm}
Very schematically the structures of the $u$- and $d$-quarks
are shown in Fig.\ref{fig:updown}, where for simplicity
the particles $\nu_e$, {\sf Y} and $e$ are visualised, respectively,  as
symbols $O$ (in grey colour), {\sf Y} and triple-{\sf Y} (coloured 
in red or blue, depending on the polarities of their charges).

\begin{figure}[htb]
\centering \epsfysize=9cm
\includegraphics[scale=0.52]{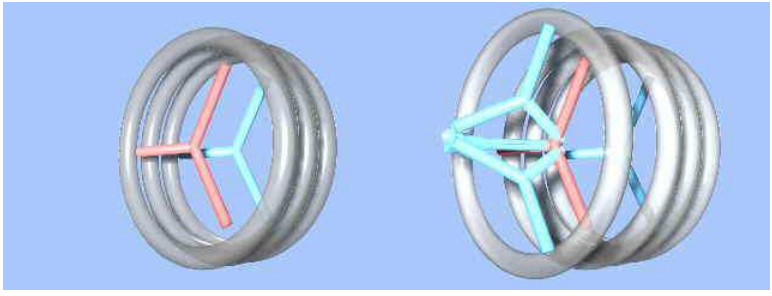}
\caption{{\small Schemes of the muon neutrino (left) and the muon (right).}}
\label{fig:numumu}
\end{figure}

\begin{figure}[htb]
\centering \epsfysize=9cm
\includegraphics[scale=0.52]{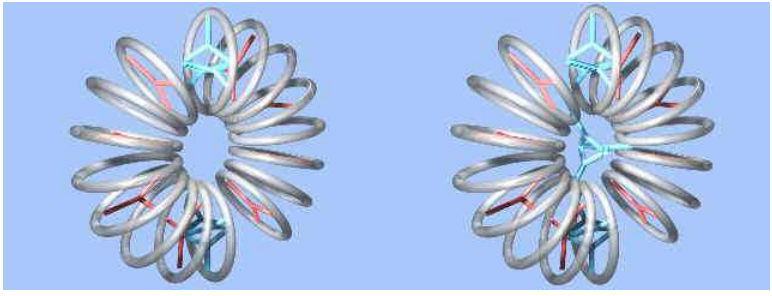}
\caption{{\small Schemes of the quarks $c$ (left) and $s$ (right).}} 
\label{fig:charmstrange}
\end{figure}

\begin{figure}[htb]
\centering \epsfysize=9cm
\includegraphics[scale=0.52]{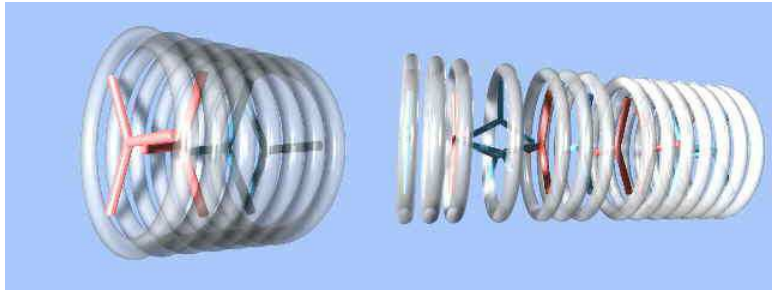}
\caption{{\small Schemes of the $\nu_\tau$ (left) and $\tau$ (right) leptons.}}
\label{fig:nutautau}
\end{figure}

\begin{figure}[htb]
\centering \epsfysize=9cm
\includegraphics[scale=0.52]{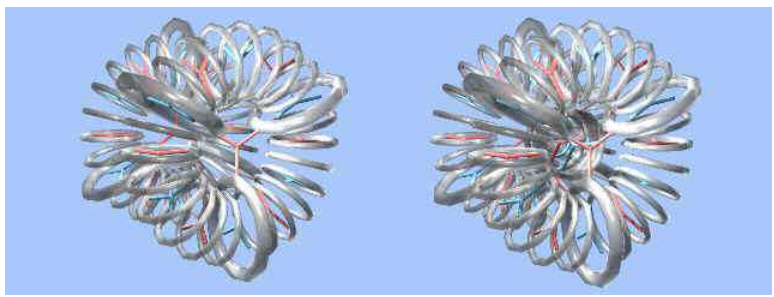}
\caption{{\small Schemes of the $t$ (left) and $b$ (right) quarks.}}
\label{fig:topbottom}
\end{figure}

\vspace{0.1cm}
It is not unnatural to suppose that particles of the second and third
generations of the fundamental fermions are formed of simpler structures 
belonging to the first generation.
For example, the muon-neutrino can be a bound state 
of the, respectively, positively and negatively charged particles 
{\sf Y$_1$} and $\overline{\sf Y}_1$ (Fig.\ref{fig:numumu}, left):

\begin{equation}
\nu_\mu= {\sf Y}\,6{\sf Y}\overline{\sf Y}_\downarrow \   6{\sf Y}\overline{\sf Y}_\uparrow
 \  6{\sf Y}\overline{\sf Y}_\downarrow\,\overline{\sf Y} 
={\sf Y}_1\nu_e\overline{\sf Y}_1, 
 \label{eq:numu}
\end{equation}
\noindent
whereas the muon's structure can be written as

\begin{equation}
\mu= (6{\sf Y}\overline{\sf Y}+3\overline{\sf Y}{\sf Y})
(6{\sf Y}\overline{\sf Y}_\downarrow\, 6{\sf Y}\overline{\sf Y}_\uparrow  
6{\sf Y}\overline{\sf Y}_\downarrow \,\overline{\sf Y})
=\overline{\nu}_e e^-\nu_\mu ,
\label{eq:muon}
\end{equation}

\noindent
which incidentally matches one of the muon's decay modes 
(Fig.\ref{fig:numumu}, right). 
The schemes of the rest of the fundamental fermions are shown in
Fig.\ref{fig:charmstrange} to Fig.\ref{fig:topbottom}. One can hardly regard these
structures as rigid bodies: they are rather oscillating clusters.
In (\ref{eq:muon}) we have enclosed  the supposedly clustered components 
in parenthesis.The oscillatory energies are likely to contribute to 
the masses of these systems. Obtaining these masses is not a straightforward task, 
but, in principle, they are computable, 
which can be shown by a simple empirical formula, relating the oscillatory energy of the 
components to the sum $M$ of their masses $m_k$:

\begin{equation}
M=\sum_k{m_k} , 
\label{eq:summass}
\end{equation}

\noindent
multiplied by the sum $\tilde{M}$ of their reciprocal masses $1/\tilde{m}_k$:

\begin{equation*}
\frac{1}{\tilde{M}}=\sum_k{\frac{1}{\tilde{m}_k}} .
\end{equation*}
 
\noindent
Then, by setting for simplicity the unit-conversion coefficient to unity,
we can compute the masses of the combined structures as

\begin{equation*}
M_{\rm total}=M \tilde{M} 
\end{equation*}

\noindent 
We shall abbreviate this summation rule by using the overlined notation: 

\begin{equation}
M_{\rm total}=\overline{m_1+ m_2+ \ldots + m_N}=\frac{m_1+m_2+\dots+m_N}
{\tilde{m}_1^{-1}+\tilde{m}_2^{-1}+\dots+\tilde{m}_N^{-1}} .
\label{eq:mtotal}
\end{equation}

\vspace{0.1cm}
The masses of the fundamental fermions computed with the use of this rule 
are summarised in Table \ref{t:massresult}. 
As an  example, let us compute the muon's mass. The masses of its 
components, according
to its structure, are: $m_1=\tilde{m}_1=48$, $m_2=\tilde{m}_2=39$, 
(in preon mass units) And the mass of the muon will be

\begin{equation*}
m_\mu =\overline{m_1+m_2} = \frac{m_1+m_2}{{\tilde{m}_1}^{-1}+
{\tilde{m}_2}^{-1}}=1872
 {\hspace{0.2cm} {\rm (preon \hspace{0.2cm} mass \hspace{0.2cm} units)}.}
\end{equation*}

\noindent
For the $\tau$-lepton, according to its structure (Fig.\ref{fig:numumu}),
 $m_1=\tilde{m}_1=156$, $m_2=\tilde{m}_2=201$, so that   

\begin{equation*}
m_\tau  = \overline{m_1+m_2}=31356 {\hspace{0.2cm} {\rm (preon 
\hspace{0.2cm}  mass \hspace{0.2cm} units)}.}
\end{equation*}

\vspace{0.1cm}
\noindent
For the proton, positively charged particle consisting of two $u$,
and one $d$ quarks submerged in a cloud of gluons $g_{ij}$, 
the masses of its components 
are $m_u=\tilde{m}_u=78$,  $m_d=\tilde{m}_d=123$,
$m_g=2m_u+m_d=279$, $\tilde{m}_g=\infty$. The resulting proton mass is 

\begin{equation}
m_p  = \overline{m_u+m_u+m_d+m_g}=
16523 {\hspace{0.2cm} {\rm (preon \hspace{0.2cm} mass \hspace{0.2cm} units)}.}
\label{eq:pmass}
\end{equation}

We can convert $m_\mu$ and other particle masses from preon mass 
units into proton mass units, $m_p$ by dividing these masses by the quantity
(\ref{eq:pmass}). The results are given  in the fourth column 
of Table \ref{t:massresult}. 
The experimental masses of the fundamental fermions 
(also expressed in units of $m_p$) are listed in the last
column for comparison 
[Groom, D.E., et al.(Particle Data Group), Eur. Phys.
Jour, C15, 1 (2000) and 2001 partial update for edition 2002
(http://pdg.lbl.gov)]

\begin{table}[htb]
\caption{{\small Computed and experimental masses of quarks and leptons.
Values given in the third column can be  converted into the proton
mass units dividing them by $m_p=16523$.}}
\label{t:massresult}
\begin{center}
\small
\begin{tabular}
{|cc|c|c|c|c|} \hline
\multicolumn{2}{|c|}{Particle and}  & {Number of }  &  Predicted    & Predicted   & Experimental \\
\multicolumn{2}{|c|}{its structure} & {preons in the}  & masses (preon & masses   & masses    \\
\multicolumn{2}{|c|}{ (components)} & {components} & mass units)   & (in $m_p$) &    (in $m_p$ units) \\ \hline
\multicolumn{6}{|c|} {First generation} \\ \hline
$\nu_e$ & 12{\sf Y}$^0$            & 36   & $\sim 0$    & $\sim 0$           & $<3\times 10^{-9}$ \\
$e^-$ & $3\overline{{\sf Y}}$               & 9    & 9    & 0.0005447 & 0.0005446170232 \\ 
$u$ & {\sf Y}$_1\nu_e${\sf Y}$_1$ & 78   & 78   & 0.00472     & 0.0021 to 0.0058 \\
$d$ & $u$ $ \ \nu_e e^-$                 & 123  & 123  & 0.007443     & 0.0058 to 0.0115 \\
\hline
\multicolumn{6}{|c|}{Second generation} \\
\hline
$\nu_\mu $ & {\sf Y}$_1$ \ \ $\nu_e$  \ \  $\overline{{\sf Y}}_1$ & 114  & $\sim 0$ & $\sim 0$ & $<2\times 10^{-4}$  \\
$\mu^-$ & \ \ $\nu_\mu$ \ \ + \   $\nu_e e^-$    & $\overline{48+39}$   & 1872 & 0.1133 & 0.1126095173 \\
$c$ & {\sf Y}$_2$ \ \ + \ \ {\sf Y}$_2$ & $\overline{165+165}$ & 27225     & 1.6477  & 1.57 to 1.95 \\
$s$ &  \ $c$ \ \ \ + \ \ $e^-$ & $\overline{165+165+9}$ & 2751 & 0.1665 & 0.11 to 0.19 \\
\hline
\multicolumn{6}{|c|}{Third generation} \\
\hline
$\nu_\tau$ & $u$ \ \ \  $\nu_e$ \ \ \  $\overline{u}$ & 192  & $\sim 0$ & $\sim 0$ & $<2\time 10^{-2}$ \\
$\tau^-$ & \ \ $\nu_\tau$ \ \  + \    $\nu_\mu \mu^-$ & $\overline{156+201}$ & 31356 & 1.8977 & $1.8939\pm 0.0003$ \\
$t$ & {\sf Y}$_3$ \  + \ \ {\sf Y}$_3$ & $\overline{1767+1767}$ & 3122289 & 188.96 & $189.7 \pm 4.5$ \\
$b$ & \  $t$ \ \ + \ \ $\mu^-$ & $\overline{1767+1767+48+39}$ & 76061.5 & 4.603 & 4.3 to 4.7  \\
\hline
\end{tabular}
\end{center}
\end{table}


\vspace{0.1cm}
Table \ref{t:massresult} and Figures \ref{fig:updown} to \ref{fig:topbottom}
illustrate family-to-family similarities in the particle
structures. In each family, the $d$-like quark is a combination of the
$u$-like quark, with a charged lepton belonging to the lighter family.
Charged leptons appear as a combination of the neutrino from the
same family with the neutrinos and charged leptons from 
the lighter family. It is conceivable that ring-closed strings, similar to that of the 
electron neutrino, may appear on higher structural levels of
the hierarchy, which could be regarded as ``heavy neutrinos'',
$\nu_h =6{\sf Y}_1{\overline{\sf Y}}_1$. They
can form part of ``ultra-heavy'' neutrinos
$\nu_{uh} =3(\overline{\sf Y}_1 \nu_h u)e^-$, and so on.
With these neutral structures the components {\sf Y}$_2$ and {\sf Y}$_3$ of the  
$c$- and $t$-quarks could be written as  
{\sf Y}$_2=u\nu_e u\nu_e e^-$ and {\sf Y}$_3= \nu_{uh}${\sf Y}.

\section*{Conclusions}

The scheme outlined here offers a reasonable explanation of the 
observed variety of matter particles. 
It generates the quantum numbers and masses of
the fundamental fermions from first principles,  
without using free or experimental input parameters 
(in this sense our model is self-consistent). The computed masses agree
with experiment to an accuracy of about 0.5\% (the discrepancies are likely 
to be attributed to the assumed simplifications, e.g., we have 
neglected the binding energies of the structures, as well as the 
small residual masses of the electron neutrino,
both contributing to the masses with opposite signs.

\end{document}